\documentclass[journal]{IEEEtran}
\ifCLASSINFOpdf
\else
\fi

\usepackage{lipsum}
\usepackage{xcolor}
\usepackage{todonotes}
\usepackage{comment}
\usepackage{arydshln}
\usepackage{cite,etoolbox}
\usepackage{times}
\usepackage{epsfig}
\usepackage{graphicx}
\usepackage{amsmath}
\usepackage{amssymb}
\usepackage[breaklinks=true,bookmarks=false]{hyperref}
\usepackage{url}
\usepackage{algorithm}
\usepackage[noend]{algpseudocode}
\usepackage{listings}
\usepackage{color}
\usepackage{setspace}
\usepackage{courier}
\usepackage{url}
\usepackage{booktabs}
\usepackage{multirow,bigdelim}
\usepackage[switch]{lineno}
\usepackage{caption}
\usepackage {afterpage}
\usepackage{xparse}

\DeclareMathOperator*{\argmax}{\arg\!\max}

\makeatletter
\NewDocumentCommand{\LeftComment}{s m}{%
	\Statex \IfBooleanF{#1}{\hspace*{\ALG@thistlm}}\(\triangleright\) #2}
\makeatother

\begin{document}
	\title{Automated Muscle Segmentation from Clinical CT using Bayesian U-Net for Personalized Musculoskeletal Modeling}

	\author{Yuta Hiasa, Yoshito Otake, Masaki Takao, Takeshi Ogawa, Nobuhiko Sugano, and Yoshinobu Sato
		\thanks{This research was supported by MEXT/JSPS KAKENHI (19H01176, 26108004), JST PRESTO
			(20407), and the AMED/ETH Strategic Japanese-Swiss Cooperative Program.}
		\thanks{(I) Y. Hiasa, Y. Otake, and Y. Sato are with Division of Information Science, Nara Institute of Science and Technology, Ikoma, Nara, Japan.}
		\thanks{(II) M. Takao is with Department of Orthopaedic Surgery, Osaka University Graduate School of Medicine, Suita, Osaka, Japan.}
		\thanks{(III) T. Ogawa, and N. Sugano are with Department of Orthopaedic Medical Engineering, Osaka University Graduate School of Medicine, Suita, Osaka, Japan.}
	}

	\markboth{}%
	{}

	\maketitle

	\begin{abstract}
		We propose a method for automatic segmentation of individual muscles from a clinical CT. The method uses Bayesian convolutional neural networks with the U-Net architecture, using Monte Carlo dropout that infers an uncertainty metric in addition to the segmentation label. We evaluated the performance of the proposed method using two data sets: 20 fully annotated CTs of the hip and thigh regions and 18 partially annotated CTs that are publicly available from The Cancer Imaging Archive (TCIA) database. The experiments showed a Dice coefficient (DC) of 0.891$\pm$0.016 (mean$\pm$std) and an average symmetric surface distance (ASD) of 0.994$\pm$0.230 mm over 19 muscles in the set of 20 CTs. These results were statistically significant improvements compared to the state-of-the-art hierarchical multi-atlas method which resulted in 0.845$\pm$0.031 DC and 1.556$\pm$0.444 mm ASD. We evaluated validity of the uncertainty metric in the multi-class organ segmentation problem and demonstrated a correlation between the pixels with high uncertainty and the segmentation failure. One application of the uncertainty metric in active-learning is demonstrated, and the proposed query pixel selection method considerably  reduced the manual annotation cost for expanding the training data set. The proposed method allows an accurate patient-specific analysis of individual muscle shapes in a clinical routine. This would open up various applications including personalization of biomechanical simulation and quantitative evaluation of muscle atrophy.
	\end{abstract}


	\IEEEpeerreviewmaketitle

	\begin{figure*}[!hbt]
		\centering
		\includegraphics[width=0.7\textwidth]{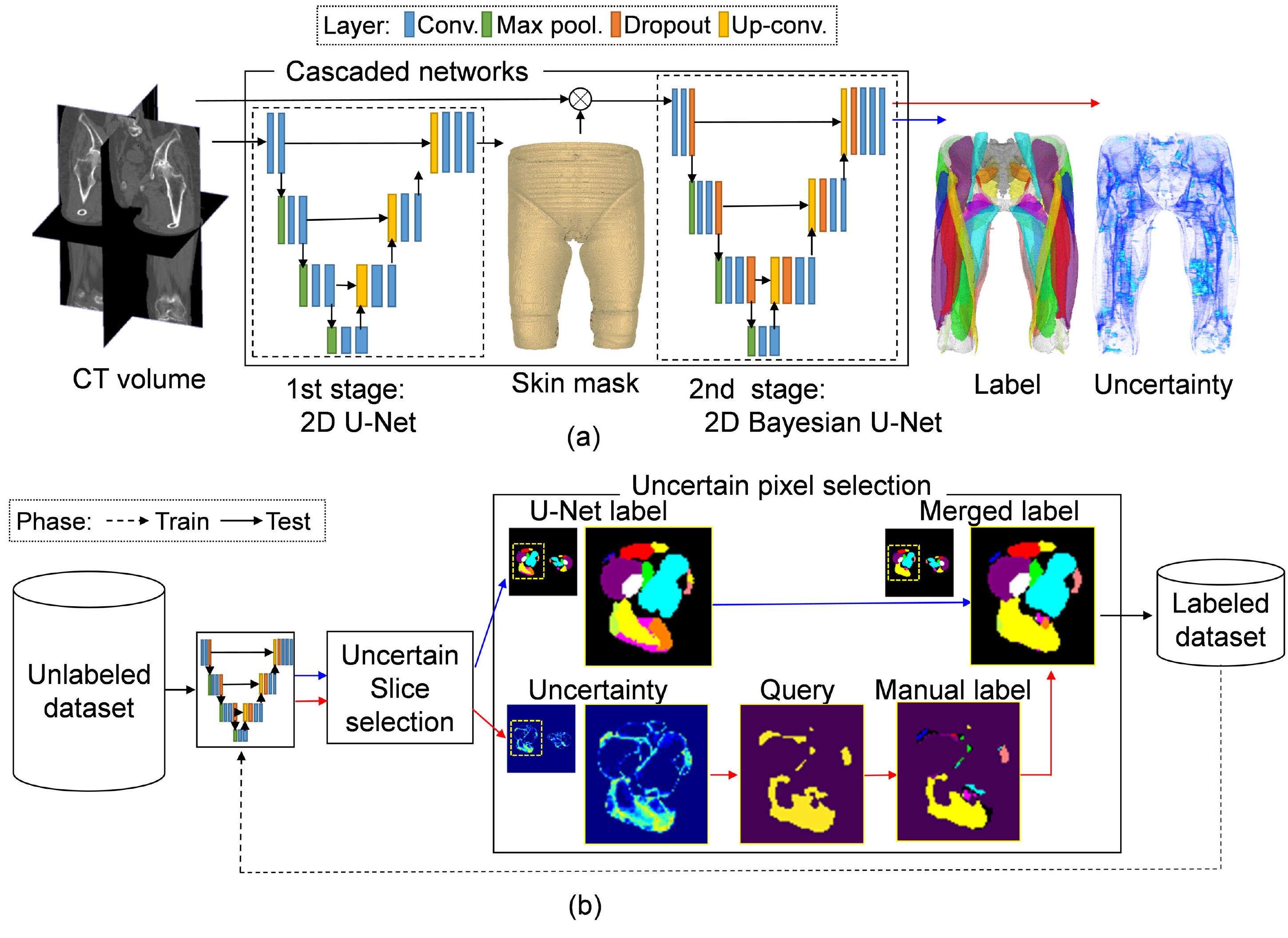}
		\caption{Workflow of the proposed methods. (a) Segmentation and uncertainty
			estimation. The skin surface is first segmented by the deterministic U-Net. Subsequently,
			the individual muscles are segmented and the model uncertainty is
			predicted by Bayesian U-Net. (b) Active-learning method.
			First, the segmentation	and uncertainty are predicted from the unlabeled images using the proposed methods. The pixels with high uncertainty are queried to the experts and relabeled manually while the pixels with low uncertainty are used directly as the training data set in the next iteration. See texts for details.}
		\label{fig:overview}
	\end{figure*}

	\section{Introduction} \label{intro}
	The patient-specific geometry of skeletal muscles plays an important role in
	biomechanical modeling. The computational simulation of human motion using
	musculoskeletal modeling has been performed in a number of studies to
	investigate musculo-tendon forces and joint contact forces, which cannot be
	easily achieved by physical measurements
	\cite{rajagopal2016full,seth2018opensim,shu2018subject}. Recent studies have
	demonstrated that personalization of model parameters, such as the size of
	the bones, geometry of the muscles and tendons, and physical
	properties of the muscle-tendon complex, improves accuracy of the simulation \cite{moissenet2017biomech, cheze2015state, taddei2012femoral}.
	While the majority of previous studies modeled the musculo-tendon unit as one
	or multiple lines joining their origin and insertion, including so-called {\it{via points}} in
	some
	cases, several recent studies have shown that volumetric models representing
	subject-specific muscle geometry provide higher accuracy in the simulation
	\cite{webb20143d}. However, the segmentation of the volumetric geometry of
	individual muscles from subject-specific medical images remains a time
	consuming task that requires expert-knowledge, thus precludes application in
	clinical practice. Therefore, our focus in this study is to develop an automated method of segmentation of
	individual muscles for personalization of the musculoskeletal model.

	\subsection{Related work} \label{subsec:related_work}
	Segmentation of muscle tissue and fat tissue has been studied extensively
	for
	the analysis of muscle/fat composition. (Note that we refer to muscle tissue
	here as an object including all muscles, not an individual muscle.) Ulbrich et
	al. \cite{ulbrich2018whole} and Karlsson et al.
	\cite{karlsson2015automatic} implemented an algorithm for automated
	segmentation of the muscle and fat tissues from MRI using a multi-atlas
	method \cite{iglesias2015multiatlas}. Lee et al. \cite{lee2017pixel} used deep
	learning for segmentation of the muscle and fat tissues in a 2D
	abdominal CT slice.

	Segmentation of individual muscles is a much more difficult problem due to the
	low tissue contrast at the border between neighboring muscles, especially in
	the area where many muscles are contiguously packed such as in the hip and thigh
	regions. Handsfield et al. \cite{handsfield2014relationships} manually
	performed segmentation of 35 individual muscles from MRIs of the lower leg in
	order to
	investigate the relationship between muscle volume and height or weight.
	To facilitate automation of the individual muscular segmentation, prior knowledge about the shape of each muscle has been introduced \cite{baudin2012prior}.
	Andrews et al. \cite{andrews2015generalized} proposed an automated segmentation
	method for 11 thigh muscles from MRI using a probabilistic shape representation
	and adjacency information. They evaluated the method using images of the middle
	part of the left femur (20 cm in length until just above the knee) and reported an
	accuracy of 0.808 average Dice coefficient. Since the muscles of interest run along a long bone, i.e., the femur, the muscles have similar appearances in axial slices resulting
	in less complexity in segmentation compared to the hip region.

	In CT images, due to the lower soft tissue contrast compared to MRI,
	segmentation of individual muscles is even more difficult. Yokota et al.
	\cite{yokota2018automated} addressed the automated segmentation of individual
	muscles from CTs of the hip and thigh regions. The target region was broader
	than \cite{andrews2015generalized} covering the origin to insertion of
	19 muscles. They introduced
	a hierarchization of the multi-atlas segmentation method such that the target
	region becomes gradually more complex in a hierarchical manner, namely starting with skin
	surface, then all muscle tissues as one object, and finally individual muscles at each
	hierarchy. They reported an average Dice coefficient of 0.838. Although their algorithm produced a
	reasonable accuracy for this highly challenging problem, due to the large number
	of non-rigid registrations required in the multi-atlas method, computational
	load was prohibitive when considering routine clinical applications (41
	minutes for segmentation of one CT volume using a high performance server with
	60 cores).

	In order to enhance the accuracy and speed of the muscle segmentation in CT, we propose an application of convolutional neural networks (CNNs). We investigate the
	segmentation accuracy as well as a metric indicating uncertainty of the
	segmentation using the framework of Bayesian deep learning. Yarin Gal et al.
	\cite{gal2016dropout} found that the dropout \cite{srivastava2014dropout} is
	equivalent to approximating the Bayesian inference, which allows estimation of
	the model uncertainty. It measures the degree of difference of each test
	sample from the training data set, originated from the deficiency of training data, namely {\it{epistemic}} uncertainty \cite{kendall2017uncertainties}. This method has been applied to brain lesion
	segmentation \cite{nair2018exploring,eaton2018towards} and surgical tool
	segmentation \cite{hiasa2018laparo}. Two example applications of the uncertainty
	metric explored in this study are; 1) prediction of segmentation accuracy without
	using the ground truth similar to the goal of Valindria et al.
	\cite{valindria2017reverse} and, 2) the active-learning framework
	\cite{maier2016crowd,yang2017suggestive} for the reduction of manual annotation
	costs.

	\subsection{Contributions}
	In this study, we demonstrate a significantly improved accuracy in the segmentation of 19 individual muscles from CTs of the hip and thigh
	regions through application of CNNs. Contribution of
	this paper is two-fold; 1) investigation of the performance of Bayesian U-Net
	using 20 fully annotated clinical CTs and 18 partially annotated CTs that are
	publicly available from The Cancer Imaging Archive (TCIA) database, 2)
	analysis of the uncertainty metric in a multi-class organ segmentation problem and
	its potential applications in predicting segmentation accuracy, without using
	the ground truth, and efficient selection of manual annotation samples in an
	active-learning framework.

	\subsection{Paper organization}
	The paper is organized as follows. In Section \ref{sec:method}, the
	proposed method is described, including data sets, uncertainty estimates, and
	active learning. In Section \ref{sec:result}, we quantitatively evaluate
	the proposed methods through experiments using two data sets. Then, we discuss the methods and results, and conclude the paper in Section \ref{sec:discuss_and_conclusion}.

	\section{Methodology} \label{sec:method}
	\subsection{Overview} \label{subsec:overview}
	Figure \ref{fig:overview} shows the workflow of the proposed methods.
	We first segment the skin surface using a 2D U-Net to isolate the body from
	surrounding objects such as the scan table and the calibration phantom. Next, the individual muscles are
	segmented and the model uncertainty is predicted using Bayesian U-Net,
	which is described in Section \ref{subsec:uncertain}. The Dice coefficient of each muscle segmentation is predicted from the model uncertainty without using the ground truth. This is done using a linear regression between the average model uncertainty computed in a cross
	validation within the training data set (Fig. \ref{fig:vis_nih}a).
	We evaluated the proposed active-learning framework, shown in Fig.
	\ref{fig:overview}(b), on a simulated environment
	using a fully annotated data set by assuming a situation where partial manual annotation is provided initially. The manual annotation of a small number of slices selected by the
	proposed procedure is given in steps as described in Section \ref{subsec:activelearn}.

	\subsection{Data sets}
	Two data sets were used to evaluate the proposed method: 1) a fully annotated non-public clinical CT data set and 2) a partially annotated publicly available CT data set.
	\subsubsection{Osaka University Hospital THA data set (THA data set)}
\begin{figure}[!bt]
		\centering
		\includegraphics[width=0.5\textwidth]{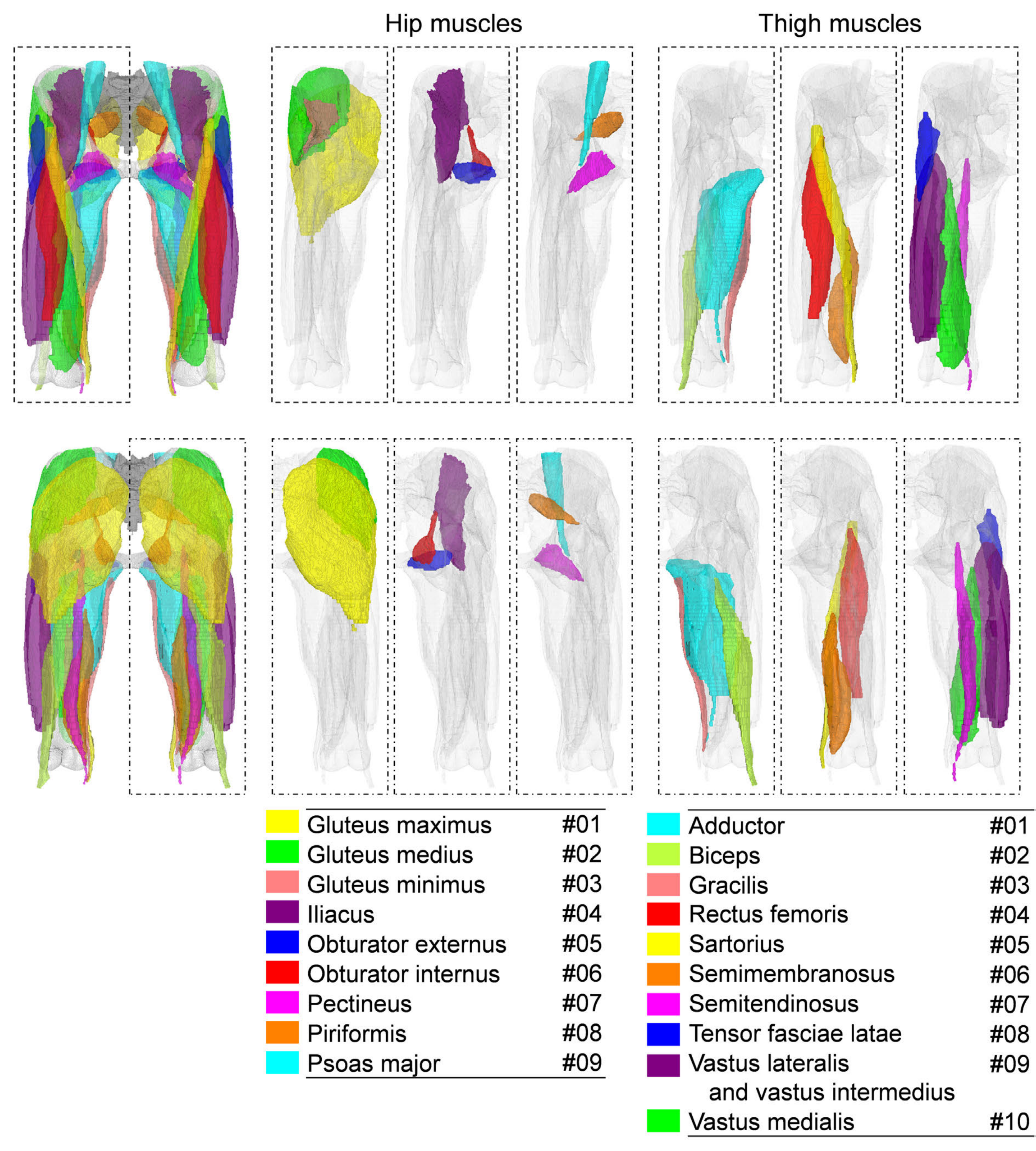}
		\caption{Training data set used in this study, consisting of 20 labeled CT volumes. The muscles of interest are separately visualized according to the functional group and their region. Upper and lower rows show the anterior and posterior views, respectively.
		}
		\label{fig:dataset}
	\end{figure}

	This data set consists of 20 CT volumes scanned at Osaka University Hospital, Suita, Japan, for CT-based planning and navigation of total hip arthroplasty (THA) \cite{yokota2018automated,ogawa2019valid}. The
	field of view was 360$\times$360 mm$^2$ and the matrix size was 512$\times$512. The
	original slice intervals were 2.0 mm for the region including the pelvis and proximal
	femur, 6.0 mm for the femoral shaft region, and 1.0 mm for the distal femur
	region.
	Each CT volume had about 500 slices (see supplementary materials for details of the number of axial slices of each muscle).
	In this study, the CT volumes were resampled so that the slice
	interval becomes 1.0 mm throughout the entire volume. Nineteen muscles around the
	hip and thigh regions and 3 bones (pelvis, femur, sacrum) were manually
	annotated by an expert surgeon (Figure \ref{fig:dataset}). The manual annotation took about
	40 hours per volume. This data set was used for training and cross-validation
	for the accuracy evaluation and prediction of the Dice coefficient.
	Note that 132 CT volumes acquired at Osaka University independently from the above mentioned data set were used for training of the skin segmentation network. The region inside the skin was semi-automatically annotated.

	\subsubsection{TCIA soft tissue sarcoma data set (TCIA data set)}
	The data set obtained from TCIA collections \footnote{\url{http://www.cancerimagingarchive.net}} contains CT and MR volumes from 51 patients with soft tissue sarcomas (STSs)
	\cite{vallieres2015radiomics}.
	In this study, we selected 18 CT volumes that include the hip region. The CT volumes were resampled so that the in-plane field of view becomes
	360$\times$360 mm$^2$ without changing the slice center and the slice interval becomes 1.0 mm
	throughout the volume similar to the THA data set. The gluteus medius muscle was manually traced by a computer scientist and verified by an expert surgeon. This data set was not used
	in the training nor in the parameter tuning and only used for evaluation of
	generalizability of the model trained with the THA data set (see Section
	\ref{subsubsec:app} for details).

	\subsection{Estimation of uncertainty metric} \label{subsec:uncertain}

	The underlying algorithm of the proposed uncertainty estimates follows that of Gal
	et al. {\cite{gal2016dropout}} which used the dropout at the inference phase. This allowed approximation of the posterior distribution based on the
	probabilistic softmax output obtained from the stochastic dropout sampling.
	We use the mean and variance of the output from multiple samplings as the
	segmentation result and uncertainty estimate, respectively.
	Below, we briefly summarize the theoretical background described in
	{\cite{gal2016dropout}}, formulate the specific metric that we employed in this
	paper, and propose a new structure-wise uncertainty metric for a multi-class
	segmentation problem.

	Suppose we have a training data set of images $\mathbf{X}=\{ \mathbf{x_1}, \cdots, \mathbf{x_n} \}$ and its labels $\mathbf{Y}=\{ \mathbf{y_1}, \cdots, \mathbf{y_n} \}$. We consider the predictive label $\mathbf{y}^*$ of an unseen image $\mathbf{x}^*$.
	Let a "{\it{deterministic}}" neural  network represent $p(y^*|\mathbf{x}^*) =
	\mathrm{Softmax}(\mathbf{f}(\mathbf{x}^*; \mathbf{W})) $. A
	"{\it{probabilistic}}" Bayesian neural network is given by marginalization over
	the weight $\mathbf{W}$ as
	\begin{eqnarray} \label{eq:bayes}
	p(y^*=c|\mathbf{x}^*, \mathbf{X}, \mathbf{Y}) = \int p(y^*=c|\mathbf{x}^*, \mathbf{W}) p(\mathbf{W} | \mathbf{X}, \mathbf{Y}) \mathrm{d}\mathbf{W}
	\end{eqnarray}
	where  $y^* \in \mathbf{y}^*$ is the output label of a pixel, $c$ is the label class, and $p(\mathbf{W} | \mathbf{X}, \mathbf{Y})$ is the posterior distribution.
	Gal et al. \cite{gal2016dropout} proved that approximation of the posterior
	distribution is equivalent to the dropout masked distribution $q(\hat{\mathbf{W}})$,
	where $\hat{\mathbf{W}} = \mathbf{W} \cdot \mathrm{diag}(\mathbf{z})$ and $\mathbf{z}
	\sim \mathrm{Bernoulli}(\theta)$, and $\theta$ is the dropout ratio. Then, Eq. (\ref{eq:bayes}) can be
	approximated by minimizing the Kullback-Leibler (KL) divergence
	$\mathrm{KL}(q(\hat{\mathbf{W}}) || p(\hat{\mathbf{W}}|\mathbf{X}, \mathbf{Y}))$ as follows.
	\begin{eqnarray}
	p(y^*=c|\mathbf{x}^*, \mathbf{X}, \mathbf{Y}) &\approx& \int p(y^*=c|\mathbf{x}^*, \hat{\mathbf{W}}) q(\hat{\mathbf{W}}) \mathrm{d}\hat{\mathbf{W}} \\
	&\approx& \frac{1}{T} \sum_{t=1}^{T} \mathrm{Softmax}(\mathbf{f}(\mathbf{x}^*, \hat{\mathbf{W}})).
	\end{eqnarray}
	where $T$ is the number of dropout samplings. This Monte Carlo
	estimation is called "{\it{MC dropout}}" \cite{gal2016dropout}. We employed the
	predictive variance as the metric indicating uncertainty which is defined as
	\begin{eqnarray}
	\lefteqn{
		Var(y^*=c|\mathbf{x}^*, \mathbf{X}, \mathbf{Y})
	} \nonumber \\
	&\approx& \frac{1}{T} \sum_{t=1}^{T} \mathrm{Softmax}(\mathbf{f}(\mathbf{x}^*, \hat{\mathbf{W}}))^T \mathrm{Softmax}(\mathbf{f}(\mathbf{x}^*, \hat{\mathbf{W}})) \nonumber \\
	& & - p(y^*|\mathbf{x}^*, \mathbf{X}, \mathbf{Y})^T p(y^*|\mathbf{x}^*, \mathbf{X}, \mathbf{Y}).
	\end{eqnarray}

	In this paper, we propose two new structure-wise uncertainty metrics: 1) predictive
	structure-wise variance (PSV) and 2) predictive Dice coefficient (PDC). PSV
	represents the predictive variance per unit area of the pixels that are
	classified as the target structure.
	Let $\mathbf{s}^*$ be all pixels that are classified as class $c$; $\mathbf{s}^*= \{y^*_i| \argmax_k p(y^*_i=k)=c, \forall y^*_i \in \mathbf{y}^* \}$ ($argmax$ represents the selection of the class with the highest probability for the pixel $i$). 
	The metric is defined as
	\begin{eqnarray}
	PSV(\mathbf{s}^*|\mathbf{x}^*) = \frac{1}{|\mathbf{s}^*|} \sum_{y^* \in \mathbf{s}^*} \sum_k Var(y^*=k|\mathbf{x}^*).
	\end{eqnarray}
	PDC is computed by a linear
	regression of PSV and the actual Dice coefficient of the target structure.
	\begin{eqnarray} \label{eq:pdc}
	PDC(\mathbf{s}^*|\mathbf{x}^*) \approx \alpha \cdot PSV(\mathbf{s}^*|\mathbf{x}^*) + \beta
	\end{eqnarray}
	where $\alpha$ is the linear coefficient and $\beta$ is the bias.
	To find these parameters, we conduct $K$-fold cross-validation. $K$-1 groups are used to train a model, while the remaining one group is used for the evaluation (i.e., observe the Dice and PSV). Then, $\alpha$ and $\beta$ are determined by all sets of observed Dice and PSV.

	As for the network architecture, we extend the U-Net model by inserting the
	dropout layer before each max pooling layer and after each up-convolution layer
	as shown in the dotted squares in Fig. \ref{fig:overview}(a), which is the same approach as
	Bayesian SegNet, proposed by Kendall et al. \cite{kendall2015bayesian}. We call
	the U-Net extended by {\it{MC dropout}} "{\it{Bayesian U-Net}}\footnote{The source code is available at {\url{https://github.com/yuta-hi/bayesian_unet}}}."

	\subsection{Bayesian active learning} \label{subsec:activelearn}
	A common practical situation in segmentation problems entails a scenario where the labeled data set is small-scale while a large-scale unlabeled data set is available. The
	active-learning method is known to be effective in that scenario by
	interactively expanding the training data set using the experts' input.

	In order to determine the pixels to query to the experts, the proposed method
	first selects slices with high uncertainty in segmentation from the unlabeled data set, which we call the {\it{slice selection step}}, and then selects pixels with high uncertainty from the selected slices, which
	we call the {\it{pixel selection step}}. The slice selection step follows Yang et al.
	\cite{yang2017suggestive} which utilized uncertainty and similarity metrics to
	determine the query slices. This is summarized as follows: Let
	$\mathcal{D}_u$ be an unlabeled data set; then a subset of uncertain slices
	$\mathcal{D}_c \subseteq \mathcal{D}_u$ is selected following the selection of
	representative slices $\mathcal{D}_r \subseteq \mathcal{D}_c$ using a similarity-based clustering approach. Details of the algorithm are shown in Appendix.

	In this paper, we propose a new method for the pixel selection step to reduce the number of pixels to query to the expert using the proposed uncertainty metric.
	We used manual labels for the pixels with uncertainty larger than the threshold $T$ (i.e., "uncertain" pixels) and predicted labels for other pixels (i.e., "certain" pixels), that is
	\begin{eqnarray} \label{eq:query}
	\hat{Y}_{ij} = \begin{cases}
	\displaystyle \argmax_k \ p(y=k|\mathbf{x}) & ( \displaystyle \sum_k Var_{ij}(y=k) < T) \\
	Y^{manual}_{ij} & (otherwise)
	\end{cases}
	\end{eqnarray}
	where $\hat{Y}_{ij}$ denotes the label for the $j$-th pixel in $i$-th slice and $Y^{manual}_{ij}$ denotes the label manually provided by the expert.
	Note that the threshold $T$ determines the trade-off between manual annotation cost and the achieved accuracy.
	We experimentally investigate the choice of the threshold $T$ in the following sections.

	\subsection{Implementation details} \label{subsec:implementation}

	During the pre-processing, intensity of the CT volumes is normalized so that [$-150,
	350$] HU is mapped to [$0, 255$] (intensities smaller than -150 HU and larger than 350 HU
	are clamped to 0 and 255, respectively). At the training phase, data
	augmentation is performed by translation of [$-25, +25$] \% of the matrix size, rotation of
	[$-10, +10$] deg, scale of [$-35, +35$] \%, shear transform with the shear
	angle of [$-\pi/8$, $+\pi/8$] rad, and flipping in the right-left direction. The data augmentation
	allows the model to be invariant to the FOV of the scan, patient's size, rotation,
	and translation. At post-processing, the largest connected component is
	extracted to obtain the final output for each muscle.

	\subsection{Comparison with conventional methods}
	The current state-of-the-art method for automated
	segmentation of individual muscles from CT based on the hierarchical
	multi-atlas method \cite{yokota2018automated} was implemented and the results were compared with the proposed method.
	In addition to U-Net, we evaluated another network
	architecture, FCN-8s \cite{long2015fully}, which is also a common fully
	convolutional neural network based on VGG16.

	We used the Dice coefficient (DC) \cite{dice1945measures} and the average symmetric
	surface distance (ASD) \cite{styner20083d} as the error metrics.
	Note that each metric was calculated per volume, not slice-by-slice.
	The statistical significance was tested by the paired $t$-test with Bonferroni correction.

	\section{Results} \label{sec:result}
	\subsection{Network architecture selection and comparison with conventional methods} \label{subsec:comarison}
	First, the segmentation accuracy is quantitatively evaluated using the 20 labeled clinical CTs, known as the THA data set. Leave-one-out cross-validation (LOOCV) was performed where a model was trained with 19 CTs and tested with the remaining one CT.
	Twenty-three class classifications (3 bones, 19 muscles, and background) were performed.
	We initialized the weights in the same way as in \cite{he2015delving}, and then
	trained the networks using adaptive moment estimation (Adam) \cite{kingma2014adam} for $1\times10^5$
	iterations at the learning rate of 0.0001. The batch-size was 3.

\begin{figure}[!bt]
		\centering
		\includegraphics[width=0.45\textwidth]{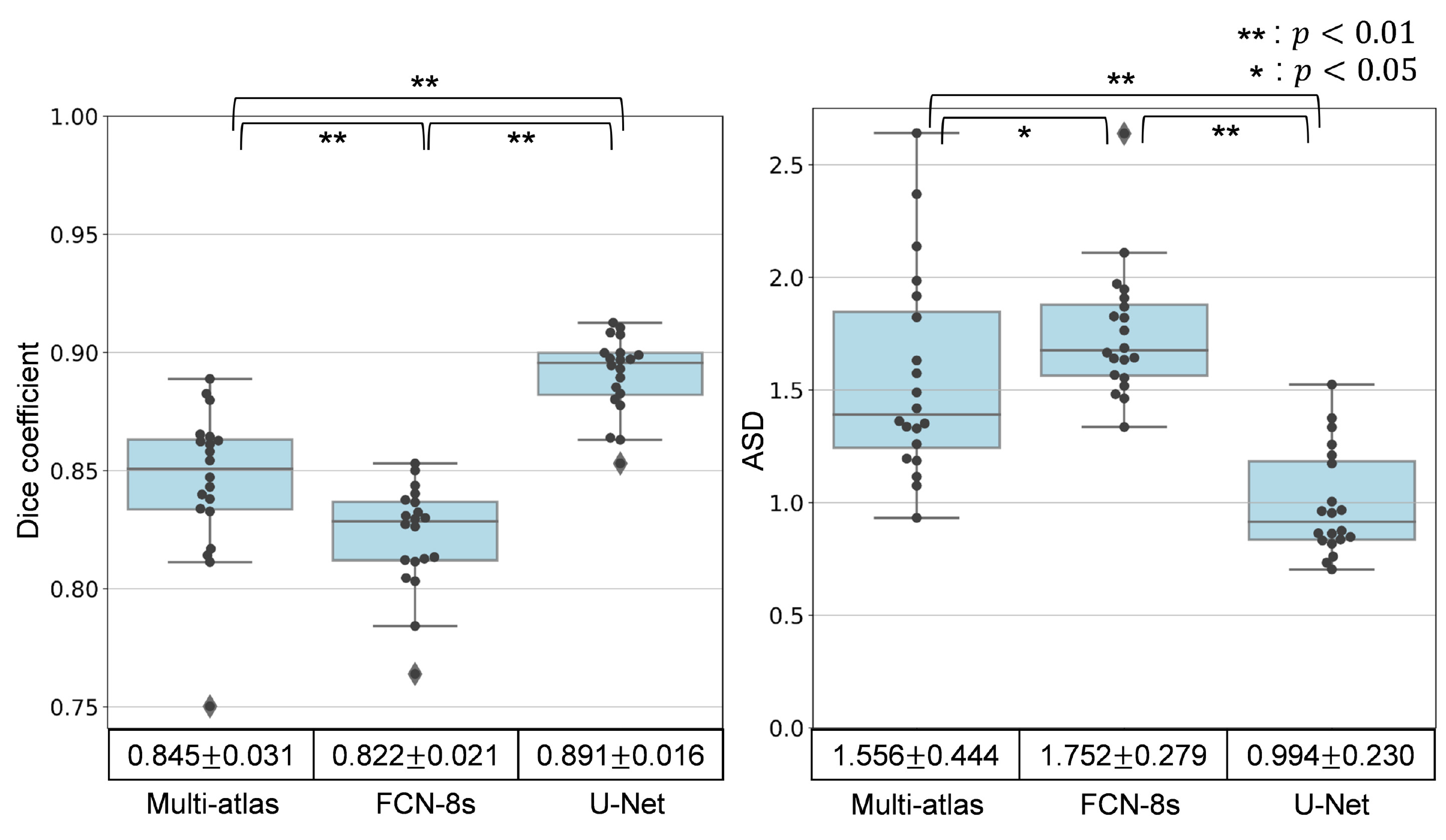}
		\caption{Accuracy of muscle segmentation for 20 patients with the hierarchical multi-atlas method \cite{yokota2018automated}, FCN-8s, and U-Net. Box and whisker plots for two error metrics: (left) Dice coefficient (DC) and (right) average symmetric surface distance (ASD). Boxes denote the 1st/3rd quartiles, the median is marked with the horizontal line in each box, and outliers are marked with diamonds. The accuracy of 19 muscles over one patient was averaged in advance (i.e., 20 data points for each box plot).}
		\label{fig:result_box}
	\end{figure}

	Figure \ref{fig:result_box} summarizes the segmentation accuracy of the muscles. The  DC and ASD over 19 muscles for one patient were averaged and plotted as box plots (i.e., 20 data points in each plot) for the multi-atlas method, FCN-8s, and U-Net. The average and standard deviation of DC for the three methods were 0.845$\pm$0.031 (mean$\pm$std), 0.822$\pm$0.021, and 0.891$\pm$0.016, respectively, while for ASD the values were 1.556$\pm$0.444 mm, 1.752$\pm$0.279 mm, and 0.994$\pm$0.230 mm, respectively. Compared with the conventional multi-atlas method \cite{yokota2018automated} and FCN-8s, U-Net resulted in statistically significant improvements ($p<0.01$) in both DC and ASD.

\begin{figure}[!bt]
		\centering
		\includegraphics[width=0.5\textwidth]{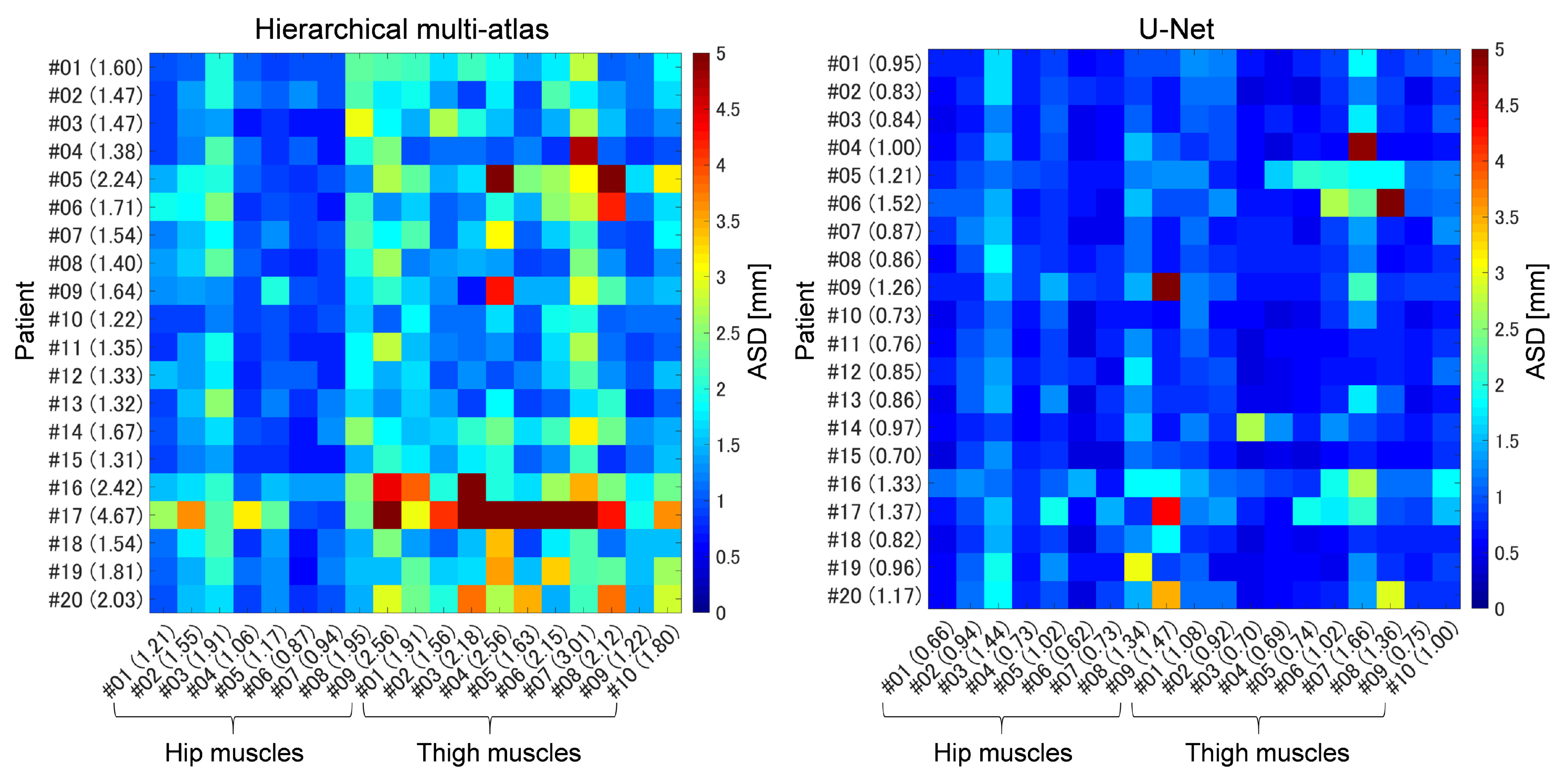}
		\caption{Heatmap visualization of ASD with  hierarchical multi-atlas method \cite{yokota2018automated} and U-Net for each individual muscle in each patient. The blue color shows higher segmentation accuracy. The numbers in parentheses indicate the mean of each row/column.}
		\label{fig:result_each}
	\end{figure}

	Figure \ref{fig:result_each} shows the heatmap visualization of ASD for the individual muscles of each patient using the multi-atlas method and U-Net. The blue color indicates a lower ASD. The accuracy improvement is clearly observed for almost all of the muscles except for 5 cells (the psoas major in Patients \#09 and \#17, gracilis in Patient \#14, semimembranosus in Patient \#04, and semimembranosus in Patient \#06).

\begin{figure}[!bt]
		\centering
		\includegraphics[width=0.5\textwidth]{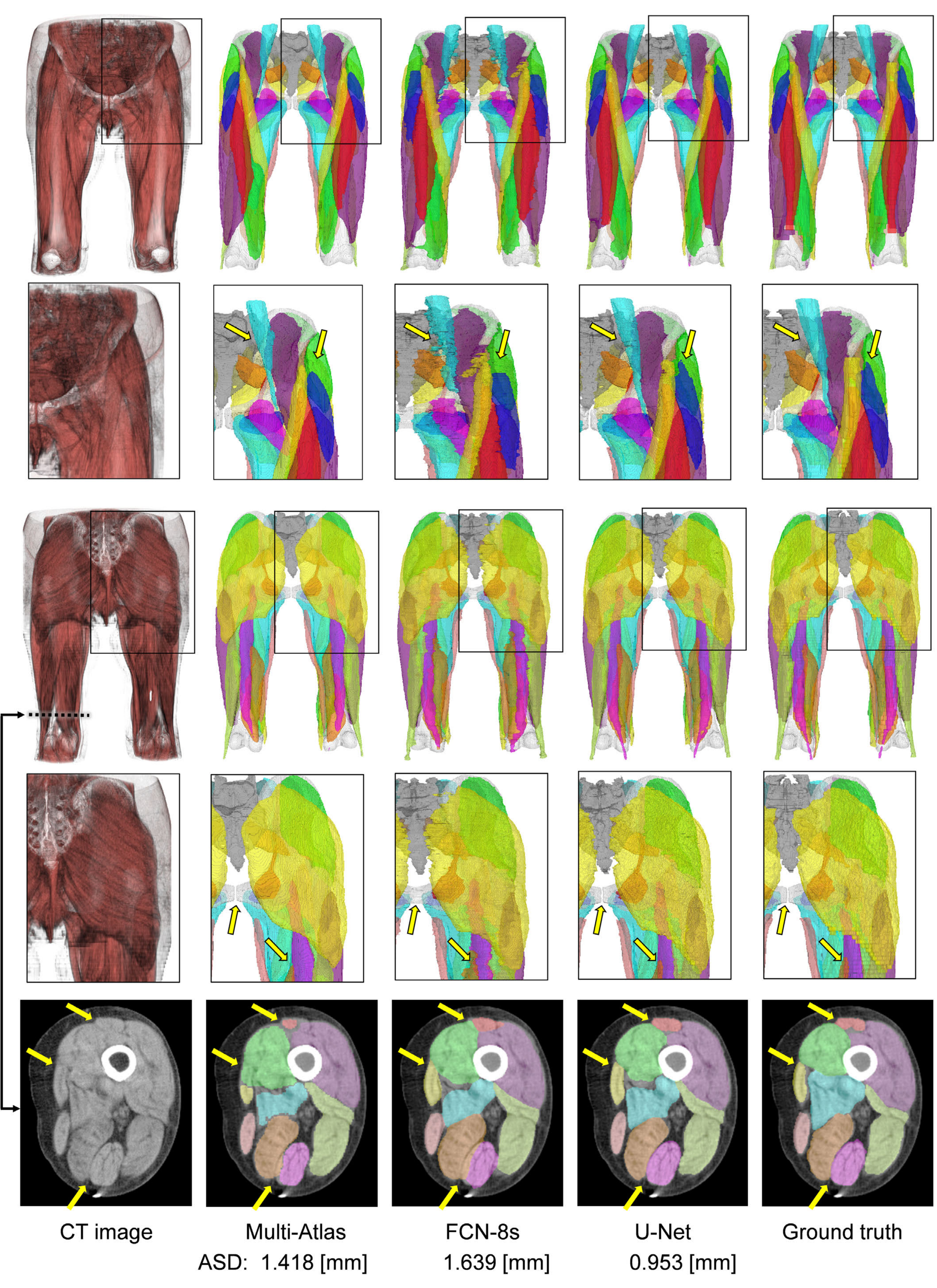}
		\caption{Visualization of the predicted label for a representative patient (Patient \#01). The result with U-Net shows distinctly more accurate segmentation near the boundary of the muscles. The region of interest in the slice visualization at the bottom corresponds to the black dotted line in the left-most column.}
		\label{fig:result_vis3d}
	\end{figure}

	Figure \ref{fig:result_vis3d} shows example visualizations of the predicted label for a representative patient (Patient \#01).
	The result with U-Net demonstrates more accurate segmentation near the boundary of the muscles compared to the other two methods. In FCN-8s, where the output layer is obtained by upsampling and fusing the latent vectors that have lower resolution (one eighth in our case) of the input size, the accuracy seemed to be consistently lower than U-Net due to the lack of details. On the other hand, in U-Net, where the output layer is directly fused with the latent vectors that have the same resolution as the input size,
	delineation of details was improved due to the pixel-wise correspondence between the input image and the output label.

	As for the segmentation accuracy of the bones with U-Net, DC of the pelvis, femur, sacrum were 0.981 $\pm$ 0.0043, 0.985 $\pm$ 0.0065, 0.962 $\pm$ 0.0166, respectively, and ASD were 0.145 $\pm$ 0.040 mm, 0.175 $\pm$ 0.084 mm, 0.402 $\pm$ 0.243 mm, respectively.

	The skin surface segmentation step did not yield statistically significant accuracy difference in the THA data set ($p>0.05$) since it did not contain such objects that added undesirable variation, but it was effective in reducing the undesirable variation for the muscle segmentation step with a low manual annotation cost, especially for the CT volumes scanned with a solid intensity calibration phantom placed near the skin surface. The calibration phantom is essential in the quantitative CT (QCT) \cite{adams2009quantitative} which is one of our main application targets for the analysis of the relationship between muscle quality and bone mineral density (see supplementary materials for evaluation of the skin surface segmentation step in QCT volumes). Note that, in our experience, simple image processing methods, such as thresholding or extraction of the largest connected component, often failed to isolate the calibration phantom from the skin surface.

	The average training time was approximately 11 hours with FCN-8s and U-Net, on
	an Intel Xeon processor (2.8 GHz, 4 cores) with NVIDIA GeForce GTX 1080Ti. The
	average computation time for the inference on one CT volume with about 500 2D slices was approximately 2 minutes excluding file loading, and
	the post-processing took about 3 minutes.

	We conducted the following experiments about the predictive accuracy and active
	learning only with the U-Net architecture, since its accuracy is significantly
	higher than the other two methods as shown above.

	\subsection{Estimation of uncertainty metric}
	\subsubsection{Relationship between uncertainty and segmentation accuracy}
	To demonstrate validity of the uncertainty metric, we investigated the relationship
	between the estimated uncertainty and the error metric using the 20 labeled CTs. We
	performed a 4-fold cross-validation where Bayesian U-Net was trained with 15
	randomly selected CTs, and tested with the remaining 5 CTs using the same
	conditions as the experiment above.

\begin{figure*}[!bt]
		\centering
		\includegraphics[width=0.95\textwidth]{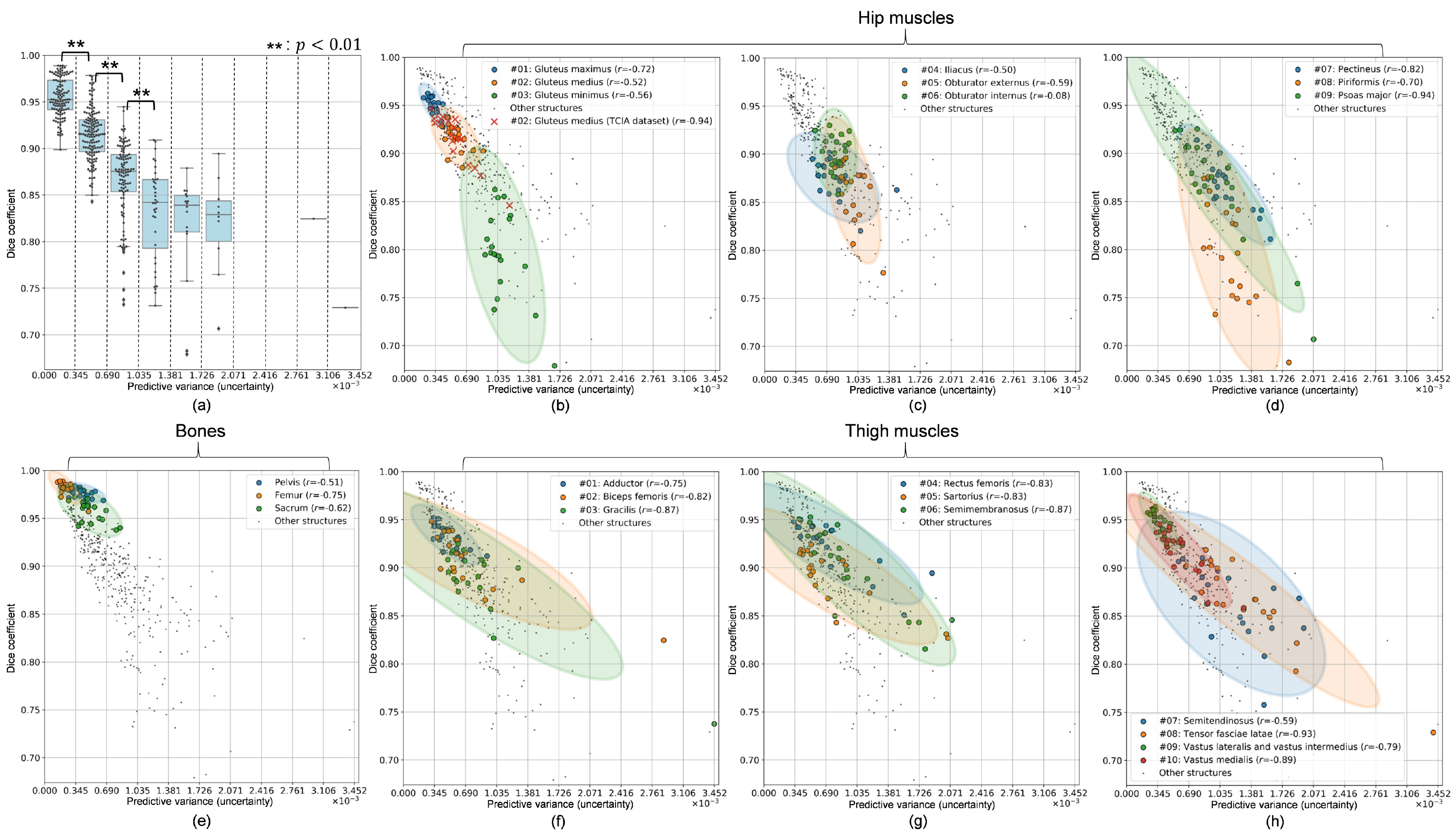}
		\caption{Relationship between the proposed uncertainty metric and segmentation accuracy. (a) Box and whisker plots of DC as a function of predictive structure-wise variance (PSV). PSV was divided into 10 bins of equal width. Mann-Whitney $U$ test was performed in adjacent bins. (b-h) Scatter plots, with the 95 \% confidence ellipses, of DC for each structure as a function of PSV. (e) Bones, and muscles of the (b-d) hip and (f-h) thigh regions. The symbol "$r$" denotes Pearson's correlation coefficient.}
		\label{fig:result_uncertain}
	\end{figure*}

\begin{figure}[!bt]
		\centering
		\includegraphics[width=0.5\textwidth]{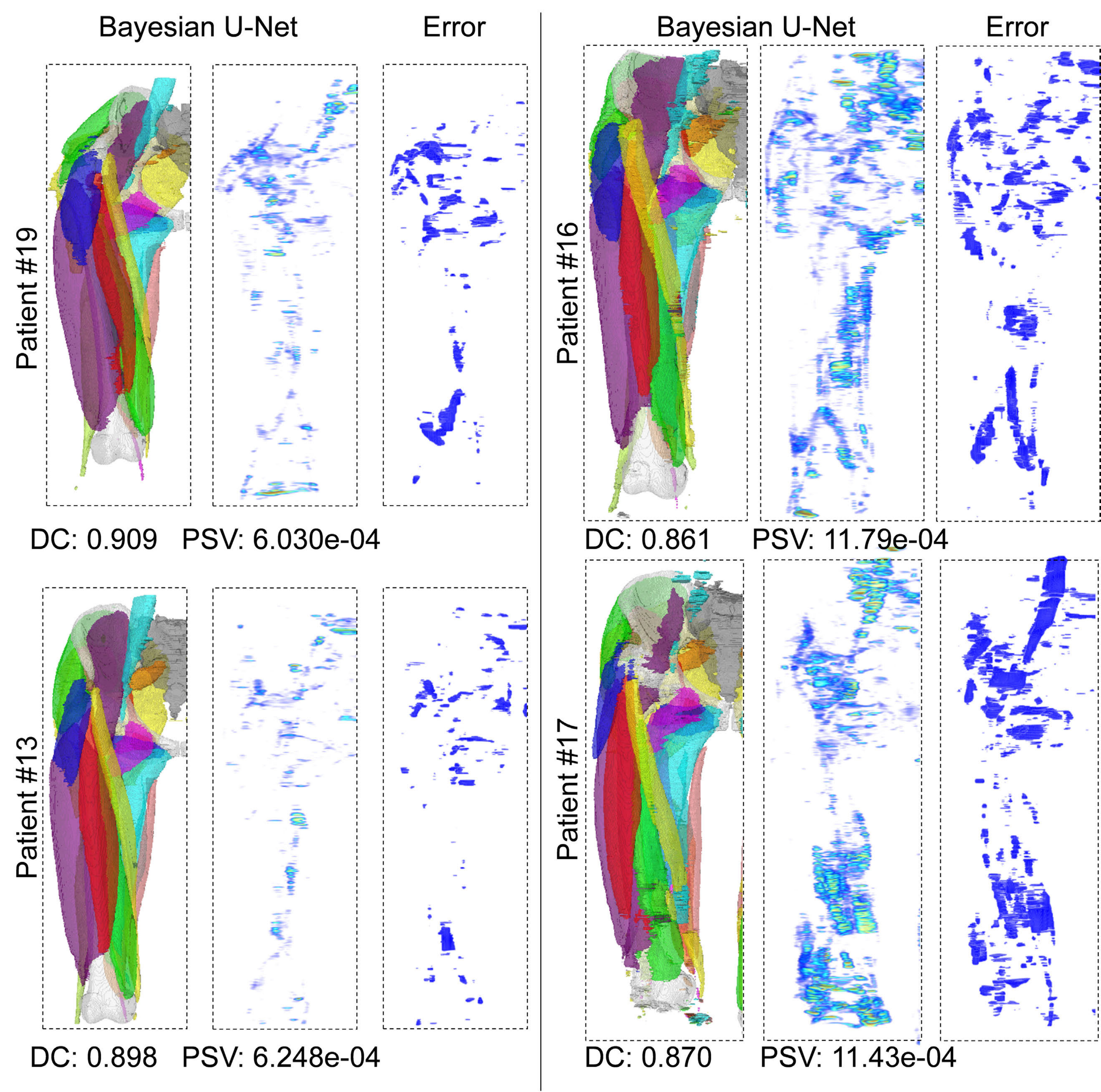}
		\caption{Visualization of the predictive variance computed by Bayesian U-Net.
			The average Dice coefficient and predictive structure-wise variance of muscles are
			denoted as DC and PSV. A good agreement between the regions with high uncertainty (denser regions in the middle sub-figure of each patient) and the regions with error (blue regions in the right sub-figure) suggests validity of the uncertainty metric to predict
			unobservable error in a real clinical situation.}
		\label{fig:vis_uncertain}
	\end{figure}

	Figure \ref{fig:result_uncertain}(a) shows the box and whisker plots of DC as a function of PSV. PSV was divided into 10 bins of equal width. The statistical significance was tested between adjacent bins, with Mann-Whitney $U$ test. The overall correlation ratio was $-0.784$.
	Figure \ref{fig:result_uncertain} (b-h) shows scatter plots of DC for the
	individual muscle structures as a function of its PSV. The 95\% confidence
	ellipses clearly illustrate the trend of the increased error (i.e., decreased DC) in accordance
	with increased uncertainty (i.e., increased PSV).
	The only muscle which had relatively low correlation was the obturator internus, which we discuss in the discussion section.
	Figure \ref{fig:vis_uncertain}
	shows an example uncertainty visualization. These high correlations between the accuracy and uncertainty suggested validity of using the uncertainty metric estimated by Bayesian U-Net as an indicator of the
	unobservable error metric without using the ground truth in a real clinical situation.

	\subsubsection{Generalization capability to an unseen data set} \label{subsubsec:app}
\begin{figure}[!bt]
		\centering
		\includegraphics[width=0.5\textwidth]{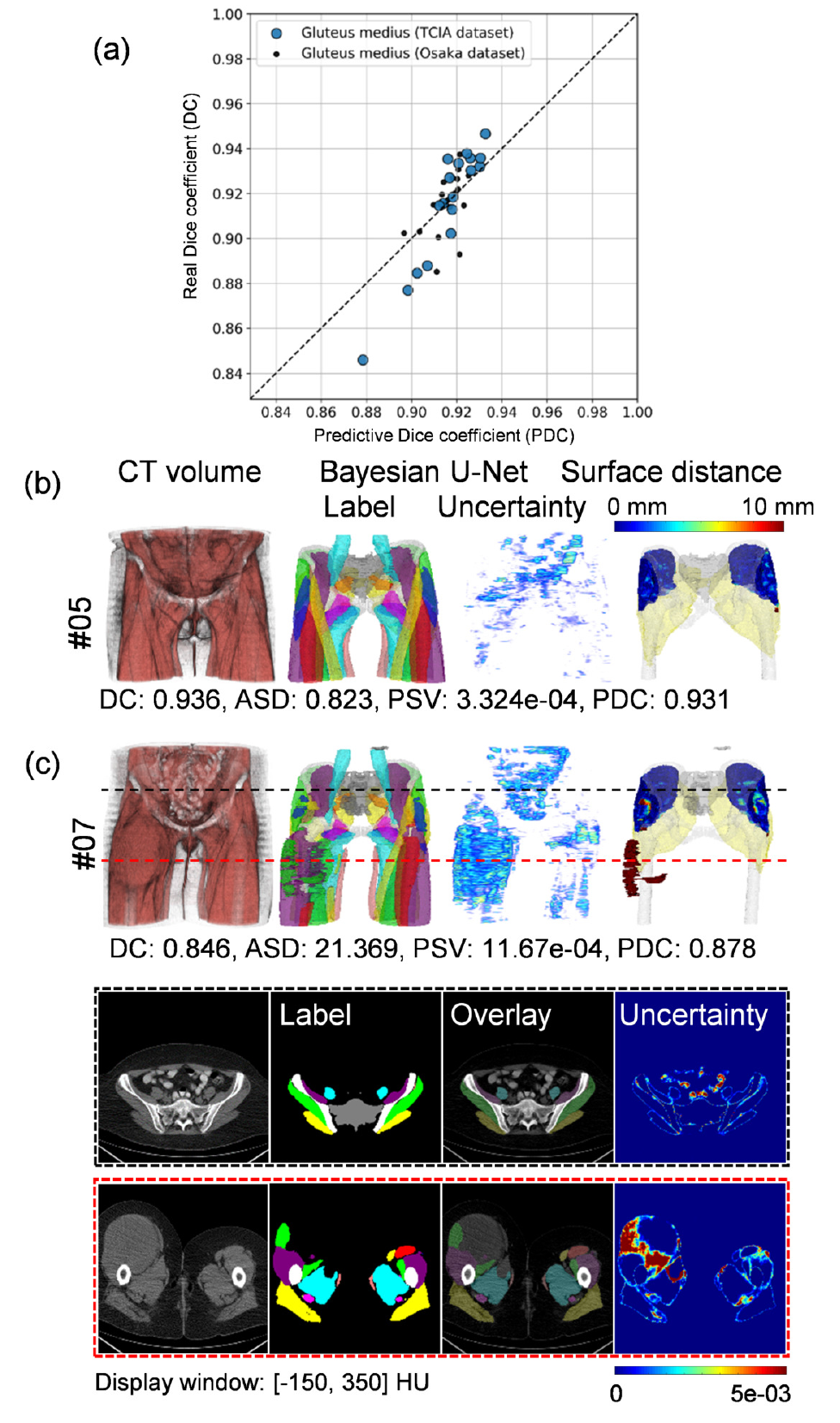}
		\caption{Evaluation of generalization capability of Bayesian U-Net on the TCIA soft tissue sarcoma data set.
			(a) Scatter plot of DC as a function of predictive Dice coefficient (PDC). (b) Representative results for one patient (\#05). (c) One patient with partial segmentation failures (\#07), from left to right: the input CT volume, the predicted label and uncertainty, and the surface distance error of the gluteus medius muscle. The predictive structure-wise variance (PSV) of the gluteus medius muscle and PDC are reported, respectively. Higher uncertainty in tumor regions was observed in Patient \#07 where the segmentation failed (shown in dark red in the surface distance error).
		}
		\label{fig:vis_nih}
	\end{figure}

	The generalization capability of Bayesian U-Net to an unseen data set was tested with
	the TCIA data set.
	Note that Bayesian U-Net was retrained using all 20 annotated CTs in the THA data set.
	Figure \ref{fig:vis_nih}(a) shows a scatter plot of DC as a function of
	PDC. $\alpha$ and $\beta$ in Eq. (\ref{eq:pdc}) were determined by
	a linear regression of 20 data points obtained from 4-fold cross-validation within the THA data set. The mean absolute error between DC
	and PDC was $0.011\pm0.0084$.
	Figures \ref{fig:vis_nih}(b) and (c) show 2 representative patients with higher and lower accuracy, respectively. The higher uncertainty regions were observed in the regions with partial segmentation failure. The quantitative evaluation in the gluteus medius muscle showed that the average DC and ASD from 18 patients were 0.914$\pm$0.026, 2.927$\pm$4.997
	mm, respectively. When excluding four outlier patients with extremely large sarcoma, the average values of DC and ASD were 0.925$\pm$0.014 and 1.135$\pm$0.777 mm, respectively, which was
	comparable to the results on the THA data set. The uncertainty was included in the plot in Fig. \ref{fig:result_uncertain}(b) (see red crosses), showing a similar distribution as the THA data set. These results suggest generalization capability of the proposed uncertainty metric between different data sets.

	\subsection{Bayesian active learning} \label{subsec:al}
	To investigate one of the application scenarios of the uncertainty estimates,
	we tested an active-learning method in a simulated environment using the 20
	fully labeled clinical CTs. The experiment assumed that 15 CTs consisting of 95\% of unlabeled slices and 5\% labeled slices were available. Then, from
	each CT, 5\% of the total number of slices from unlabeled slices was manually or automatically labeled and added
	to the labeled data in one step, which we call one "{\it{acquisition step}}."
	We iterate the acquisition step 20 times. The remaining 5 CTs were used as the
	test data set. In each acquisition step, Bayesian U-Net was initialized and
	trained using Adam \cite{kingma2014adam} for maximal $300$ epochs at the
	learning rate of 0.0001 with the early stopping schema.
	Note that each axial CT slice was downsampled to $256\times256$ in this experiment due to the limitation of training time. The data augmentation was purposely not performed in order to investigate the behavior of the model purely dependent on the number of training data sets.

	For a quantitative evaluation of the manual labor, we defined a metric that we call {\it{manual annotation cost}} (MAC) as
	\begin{eqnarray}
	MAC (Y) = \frac{|Y^{manual}|}{|Y|}
	\end{eqnarray}
	where $Y$ is the added label image. $|Y^{manual}|$ denotes the number of pixels to be queried in $Y$.

	We compared the segmentation accuracy at each acquisition step with the following
	three pixel selection methods. (1) Fully-manual selection
	\cite{yang2017suggestive}: The user annotates all pixels in the uncertain
	slices. (2) Random selection: The user annotates random pixels. (3) Semi-automatic selection (proposed method): The user annotates only
	uncertain pixels. In order to perform a fair comparison, we set the experimental condition so that the number of pixels annotated in (2)
	and (3) were equal. Note that the fully-manual selection results in $MAC=1.0$.

\begin{figure}[!bt]
		\centering
		\includegraphics[width=0.4\textwidth]{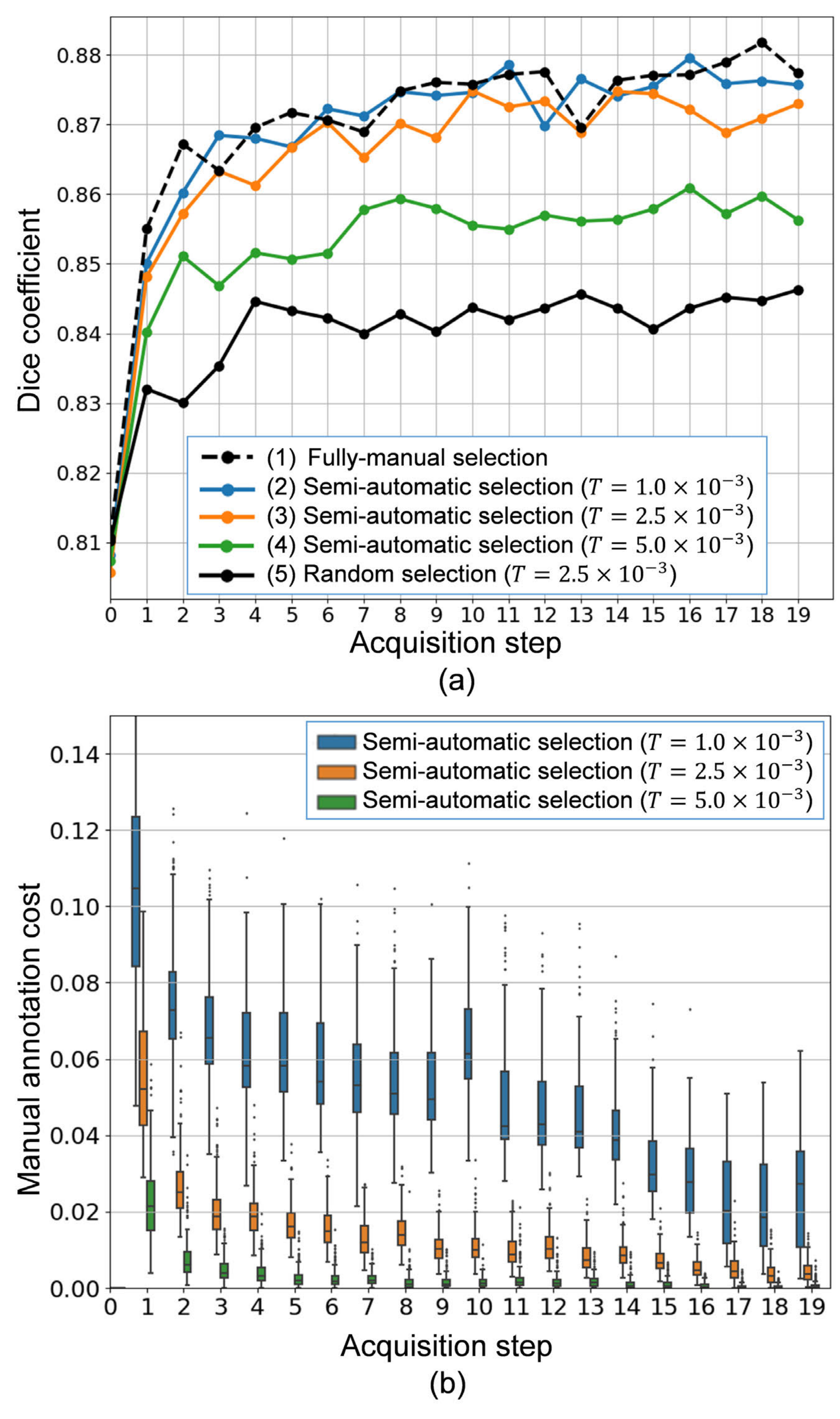}
		\caption{Results of the active-learning experiment using the proposed pixel selection method. (a) The plot of mean DC over individual structures and patients as a function of the acquisition step for different pixel selection methods. (b) The box and whisker plots of manual annotation cost at each acquisition step.}
		\label{fig:result_al_dice}
	\end{figure}

	Figure \ref{fig:result_al_dice}(a) shows mean DC over all muscles and patients as a function of the acquisition step (note that each acquisition step adds 5\% of the total training data set resulting in 100\% after 20 steps). The proposed semi-automatic selection was tested with three different uncertainty thresholds, $T$ in Eq. (\ref{eq:query}).
	For a larger $T$, we trust a larger number of pixels in the automatically estimated labels and only those pixels with highly uncertain pixels will be queried to the experts. For a smaller $T$, we trust less number of pixels in the automatically estimated labels and more pixels will be queried to the experts, resulting in a higher MAC.
	Figure \ref{fig:result_al_dice}(b) shows the MAC metric at each acquisition step.
	First, we observed a trend that the accuracy increases as the training data set increases with any selection method. The random selection method stopped the increase at around a DC of 0.843, while the other two methods kept increasing. The DC of the proposed method with $T<=2.5\times10^{-3}$ reached a DC higher than the random selection by about 0.03, which was close to the fully-manual selection method. When comparing the three thresholds in the proposed method, the larger number of pixels were queried (i.e., larger MAC) when the threshold was low; however, it did not reach the DC value achieved via fully-manual selection when the threshold was too low, i.e., $T=5.0\times10^{-3}$. MAC gradually decreases according to the acquisition step, because the overall certainty increased according to the increase of training data set. In this experiment, we concluded that the threshold with a good trade-off between achievable accuracy and annotation cost was $T=2.5\times10^{-3}$, which resulted in an approximately 90-fold cost reduction compared to fully-manual selection (i.e., median MAC was 0.0108 over all 19 acquisition steps). Note that the median MAC in case of  $T=1.0\times10^{-3}$ and $5.0\times10^{-3}$ were 0.0484 and 0.0013, respectively.

	\begin{figure}[!bt]
		\centering
		\includegraphics[width=0.5\textwidth]{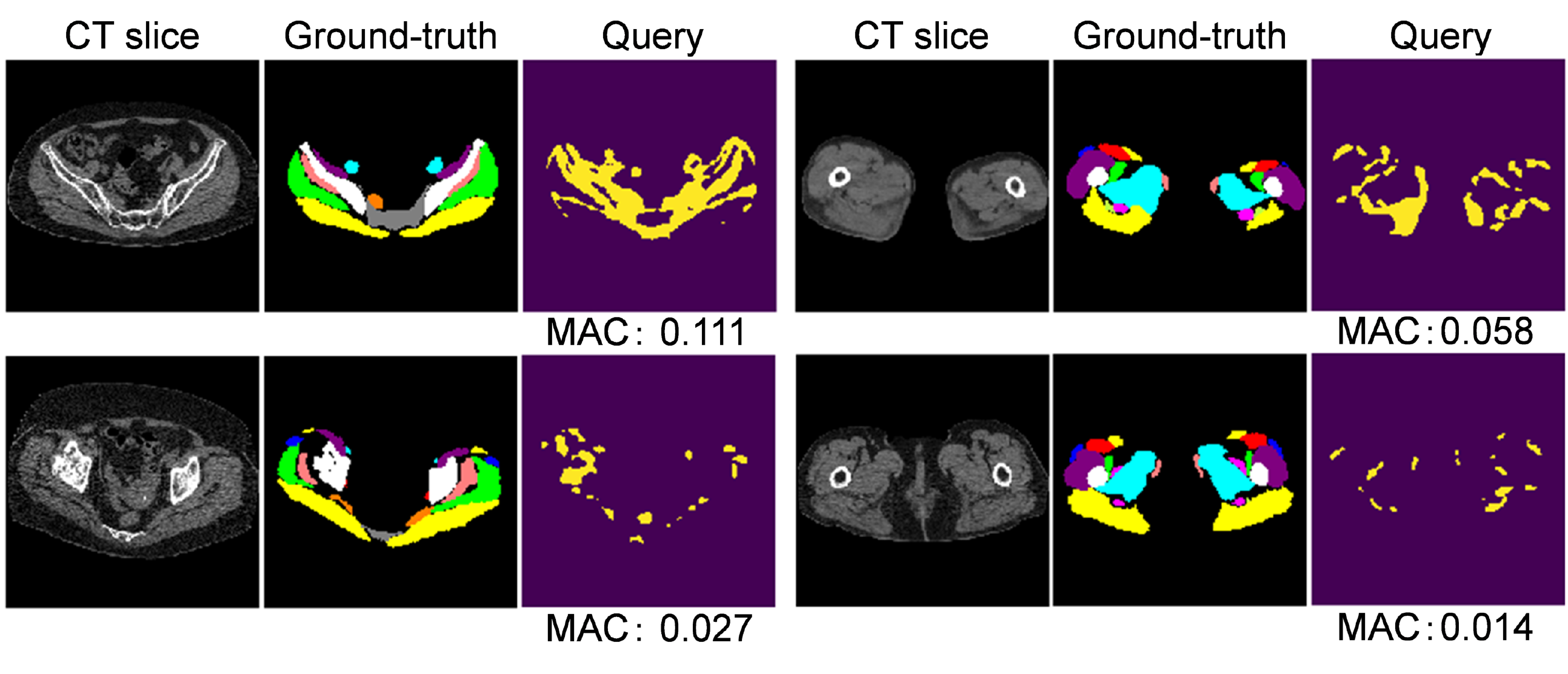}
		\caption{Examples of query pixels to be manually annotated (colored by yellow) and their manual annotation cost (MAC).}
		\label{fig:result_al_vis}
	\end{figure}

	\section{Discussion and Conclusion} \label{sec:discuss_and_conclusion}

	We presented the performance of CNNs for use in segmentation of 19 muscles in the lower extremity in clinical CT. The findings in this paper are three-fold. The proposed Bayesian U-Net 1) significantly improved segmentation accuracy over the state-of-the-art hierarchical multi-atlas method and demonstrated high generalization capability to unseen test data sets, 2) provided prediction of the quantitative accuracy measure, namely the Dice coefficient, without using the ground truth, and 3) can be used in the active-learning framework to achieve considerable reduction in manual annotation cost.

	The LOOCV using 20 fully annotated CTs showed the average DC of 0.891$\pm$0.016 and ASD of 0.994$\pm$0.230 mm, which were significant improvements ($p<0.01$) when compared with the state-of-the-art methods. The muscles that exhibited ASD larger than 3 mm with Bayesian U-Net (see Fig. \ref{fig:result_each}) were  the piriformis (hip \#08) of Patient \#19,  the psoas major (hip \#09) of Patients \#09, \#17, and \#20, the semitendinosus (thigh \#07) of Patient \#04, and the tensor fasciae latae (thigh \#08) of Patient \#06. After careful verification of those 6 muscles, we found one error in the ground truth expert's trace (thigh \#07 muscle of Patient \#04). The accuracy and inter-/intra-operator variability in the manual trace is a frequently raised question. In our case, several rounds of inspections and reviews among the expert group were performed on the manual traces, especially on some muscles, which are difficult to define their boundaries even by experts, and finally consensus among the expert group was established.
	We consider that the proposed Bayesian U-Net learned the trace generated by the experts specialized in musculoskeletal anatomy and correctly reproduced the trace that would have been created by an expert in the same group with high fidelity. The muscles with higher average ASD (hips \#03, \#08, \#09, thighs \#07, \#08) had specifically obscure boundaries in the axial plane, and an especially larger error among them was observed in muscles elongated in z- (superior-inferior) direction (hip \#09 and thigh \#07). On the other hand, the thigh muscles showed notably higher error in the multi-atlas method than Bayesian U-Net, because the thigh muscles, especially the gracilis (thigh \#03), which is a thin muscle located near the skin surface in the lower thigh region, exhibited a larger shape variation than the hip muscles due to the variation in the hip joint position. These muscles are susceptible to the error in the registration that relies on the spatial smoothness in 3D, while our 2D slice-by-slice segmentation approach was not affected.

	As for the uncertainty metric for prediction of accuracy, the high correlation between uncertainty and Dice coefficient in both THA and TCIA data sets suggested the potential for its use as the performance indicator.
	The only muscle with low correlation ($r=0.08$) was the obturator internus (hip \#06). A possible reason of the low correlation is that non-{\it{epistemic}} variability became dominant. The obturator internus is a small muscle connecting internal surface of the obturator membrane of the pelvis and medial surface of the greater trochanter of the femur and traveling almost in parallel to the axial plane (see Fig. \ref{fig:dataset}).
	We believe these properties entailed a challenge in manual tracing and the variability in the ground truth (so-called {\it{aleatory}} variability) became dominant.
	The psoas major (hip \#09) and the tensor fasciae latae (thigh \#08) had major failures in a few cases, but their low uncertainty metrics correctly indicated the failures. Valindria et al. \cite{valindria2017reverse} also attempted to predict the segmentation performance without using the ground truth by using the predicted segmentation of a new image as a surrogate ground truth for training a classifier which they call a reverse classification accuracy (RCA) classifier. They tried three different classifiers for use as segmentation and the RCA classifiers, and investigated the best combination exhaustively. Extensive comparative studies with our approach are intriguing, but it is beyond the scope of this paper. However, our approach using the MC dropout sampling representing the {\it{epistemic}} uncertainty in the model would be a more straightforward strategy to performance prediction without requiring an exhaustive search. The uncertainty metrics were recently investigated by Eaton-Rosen et al. \cite{eaton2018towards} in a binary segmentation problem of the brain tumor, specifically for quantifying the uncertainty in volume measurement. Nair et al. \cite{nair2018exploring} also explored uncertainty in binary segmentation for lesion detection in multiple sclerosis. Our present work is distinct from these previous works in that we demonstrated correlation between the structure-wise uncertainty metric, namely MC sample variance, and the Dice coefficient of each structure.

	Active learning, in which the algorithm interactively queries the user to obtain the desired ground truth for new data points, is an extensively studied topic including discussion regarding the efficient use of non-expert knowledge from the {\it{crowd}}  \cite{maier2016crowd} and the efficient savings of the manual annotation cost by the expert \cite{yang2017suggestive}. We enhanced the approach developed by Yang et al. \cite{yang2017suggestive} which selected the new image of which the expert's ground truth is most effective to improve accuracy. Our proposal is to further reduce the annotation cost by focusing on pixels to annotate, resulting in an approximately 90-fold cost reduction.
	The idea of pixel selection is similar to that proposed in \cite{maier2016crowd}, in which only super-pixels with high uncertainty is manually annotated. In summary, the proposed method combines slice- \cite{yang2017suggestive} and pixel- \cite{maier2016crowd} selection methods based on Bayesian neural networks \cite{gal2016dropout}.
	Our algorithm introduces one additional hyper parameter, which is the threshold of the uncertainty determining the pixel to be queried or not. We experimentally demonstrated that the threshold determined the trade-off between the manual annotation cost, learning speed, and final achievable accuracy. The optimum choice of the threshold value for a new data set requires further theoretical and experimental considerations, although the rate of initial improvement in accuracy during the first few acquisition steps would provide indications about the behavior in further steps as shown in Fig. \ref{fig:result_al_dice}(a).
	Stopping criteria in active learning have been discussed in \cite{settles2009active}. The ideal criterion is when the "cost" caused by the error (e.g., incorrect diagnosis) becomes less than the annotation cost. However, in practice, the "cost" caused by the error is difficult to estimate, so the active learning is usually stopped when the learning curve stalls.

	Our target application mainly focuses on personalization of the biomechanical simulation. The volumetric muscle modeling, using a finite element model \cite{webb20143d,yamamura2014effect} or a simpler approximation in shape deformation for real-time applications such as \cite{murai2016anatomographic}, has shown advantages in accurate prediction of muscle behavior. In addition, Otake et al. \cite{otake2018registration} demonstrated the potential for estimating the patient-specific muscle fiber structure from a clinical CT assuming the segmentation of each muscle was provided. The proposed accurate automated segmentation method enhances this volumetric modeling in clinical routine as well as in studies using a large-scale CT database for applications such as statistical analysis of human biomechanics for ergonomic design. The patient-specific geometry of skeletal muscles has also been studied in clinical diagnosis and monitoring of muscle atrophy or muscle fatty degeneration caused by or associated with conditions such as trauma, aging, disuse, malnutrition, and diabetes \cite{rasch2009persisting,uemura2016volume}, where muscles were delineated manually by a single operator from the images. The automated segmentation is also advantageous in the reduction of the manual labor and inter-operator variability in these analyses.

	In general, CT is superior in terms of speed compared to MRI. The CT scanning protocol that we used for the lower extremity took less than 30 seconds, while a typical MRI scan of the same range with the same spatial resolution would require more than 10 minutes. The fast scan is especially advantageous in orthopedic surgery, where biomechanical simulation is most helpful, to obtain the entire muscle shapes from their origin to insertion in the thigh region. Nonetheless, application of the proposed method to MR images would also be achievable, for example, by using an algorithm such as CycleGAN \cite{hiasa2018cross,zhang2017deep} for synthesizing a CT-like image from the MR image.

	One limitation in this study is the limited variation in the training and test data set. The THA data set only contains females who were subject to THA surgery, which limits variation in size and fat content in muscles. Although the TCIA data set contains male patients and a larger variation in terms of pathology, the ground truth label is available only for the gluteus medius muscle. Another limitation in the active-learning method is that the experiment was only a simulation. Although it illustrated potential usefulness of the proposed uncertainty metric with dependency on the uncertainty threshold in one type of active-learning framework, further investigation with a larger labeled- and unlabeled- CT database would be preferable to evaluate effectiveness of the proposed method in a more realistic clinical scenario. 
	An investigation of an effective learning algorithm that exploits information from a large-scale unlabeled data set without requiring the iterative/time-consuming manual annotation is also in our future work.

	\newpage
	\afterpage{\clearpage}

	\appendix \label{sec:suppl}
	\begin{algorithm}
		\caption{Slice selection by similarity-based clustering}
		\label{alg:slice}
		\begin{algorithmic}[1]
			\Require{unlabeled data set $\mathcal{D}_u$; uncertain slices $\mathcal{D}_c \subseteq \mathcal{D}_u$; representative slices $\mathcal{D}_r = \emptyset$}
			\While{$|\mathcal{D}_r| < N$}
			\Comment{{\footnotesize{Select $N$ representative slices}}}
			\State{$l, i_{best} \leftarrow 0, 0$}
			\For{$i \leftarrow 1$ to $|D_c|$}
			\LeftComment{{\footnotesize{Find the next best representative image from $\mathcal{D}_c$ that maximizes similarity between $\hat{\mathcal{D}}_r$ (tentative $\mathcal{D}_r$) and $\mathcal{D}_u$ }}}
			\State{$\hat{\mathcal{D}}_r, m \leftarrow \mathcal{D}_r \cup \{I_{c,i}\}, 0$}
			\Comment{{\footnotesize{$I_{c,i} \in \mathcal{D}_c$}}}
			\For{$j \leftarrow 1$ to $|\mathcal{D}_u|$}
			\LeftComment{{\footnotesize{Calculate similarity between $\hat{\mathcal{D}}_r$ and $\mathcal{D}_u$}}}
			\State{$n \leftarrow 0$}
			\For{$k \leftarrow 1$ to $|\hat{\mathcal{D}}_r|$}
			\State{$s$ $\leftarrow$ $similarity$ ($I_{u,j}$, $\hat{I}_{r,k}$)}
			\Comment{{\footnotesize{$I_{u,j} \in \mathcal{D}_u$}}}
			\If{$s>n$} {$n \leftarrow s$}
			\EndIf
			\EndFor
			\State{$m \leftarrow m+n$}
			\EndFor

			\If{$l > m$} {$l, i_{best} \leftarrow m, i$}
			\EndIf
			\EndFor
			\State{$\mathcal{D}_r \leftarrow \mathcal{D}_r \cup \{I_{c, i_{best}}\}$}
			\EndWhile
		\end{algorithmic}
	\end{algorithm}

	\ifCLASSOPTIONcaptionsoff
	\newpage
	\fi

	\clearpage

	{\small
		\bibliographystyle{ieee}
		\bibliography{refs}
	}

	\onecolumn
	\newpage
	\afterpage{\clearpage}

	\makeatletter
	\def\suppl{\relax}
	\renewcommand{\suppl}[1][]{\par%
		\def\theHsection{Suppl.A}%
		\xdef\Hy@chapapp{suppl}%
		\setcounter{section}{0}%
		\setcounter{subsection}{0}%
		\setcounter{subsubsection}{0}%
		\setcounter{paragraph}{0}%
		\def\thesection{}%
		\def\thesectiondis{}%
		\def\thesubsection{\Alph{subsection}}%
		\refstepcounter{section}
		\@ifmtarg{#1}{\@IEEEappendixsavesection*{Supplementary materials}%
			\addcontentsline{toc}{section}{Suppl.}}{%
			\@IEEEappendixsavesection*{Suppl. \\* #1}%
			\addcontentsline{toc}{section}{Suppl.: #1}}%
		\def\section{\@ifstar{\@IEEEappendixsavesection*}{%
				\@IEEEdestroythesectionargument}}
	}

	\renewcommand{\figurename}{Suppl.}
	\renewcommand{\thefigure}{\Alph{figure}}
	\setcounter{figure}{0}
	\makeatother

	\suppl
	
	\subsection*{}
		\begin{figure}[!b]
		\centering
		\includegraphics[width=0.93\textwidth]{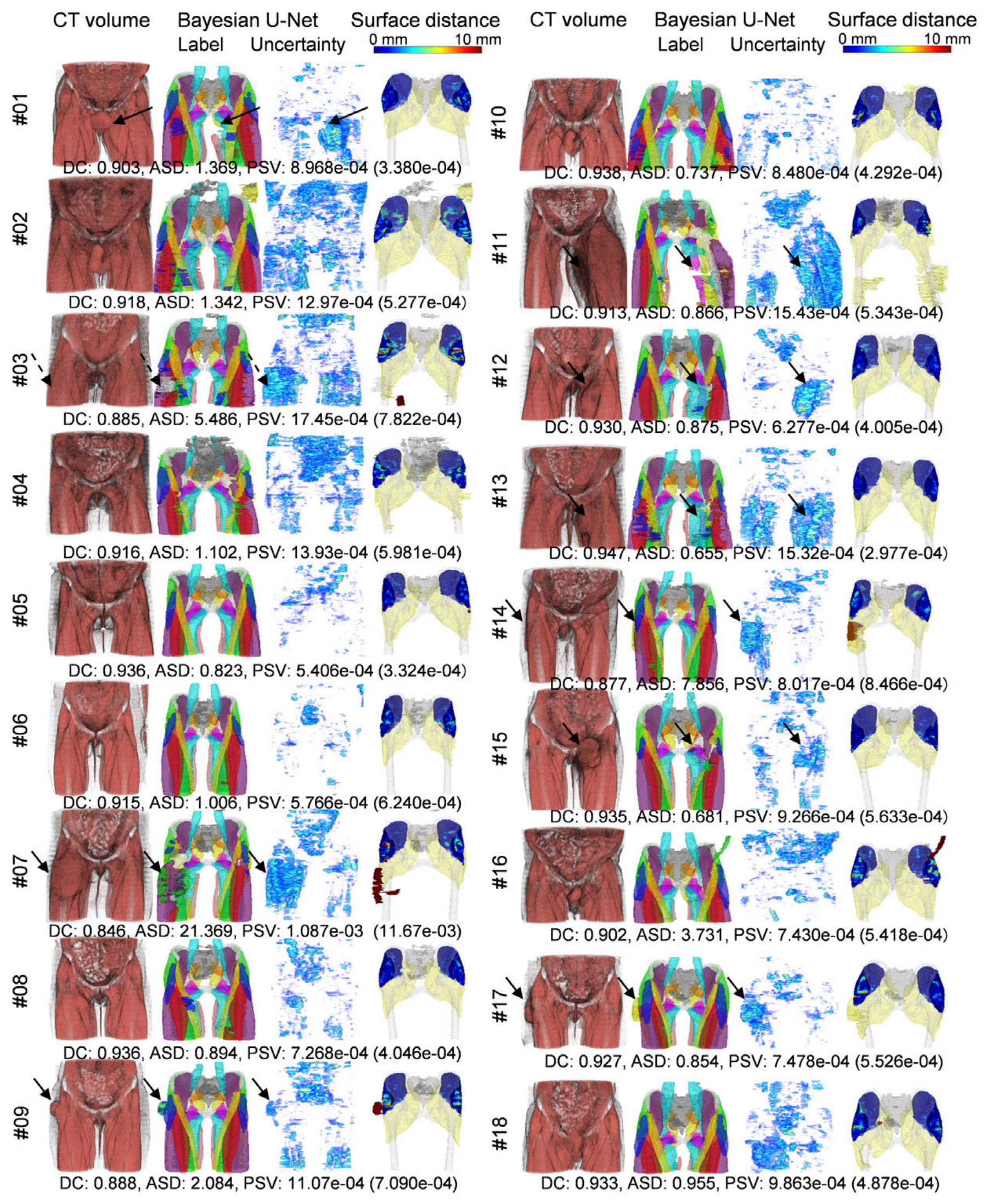}
		\caption{Bayesian U-Net on the TCIA soft tissue sarcoma data set. The tumor caused mis-segmentation in Patients \#01, \#07, \#09, \#11-\#15 and \#17, we observed that such failed regions indicated high uncertainty (black solid allow). In Patient \#03, CT artifacts led to failure (black dashed allow). In Patient \#10, some thigh muscle structures were out of the FOV, which led to mis-segmentation. The uncertainty indicated high value also in these regions.
		}
		\end{figure}

	\subsection*{}
		\begin{figure}[!b]
			\centering
			\includegraphics[width=0.9\textwidth]{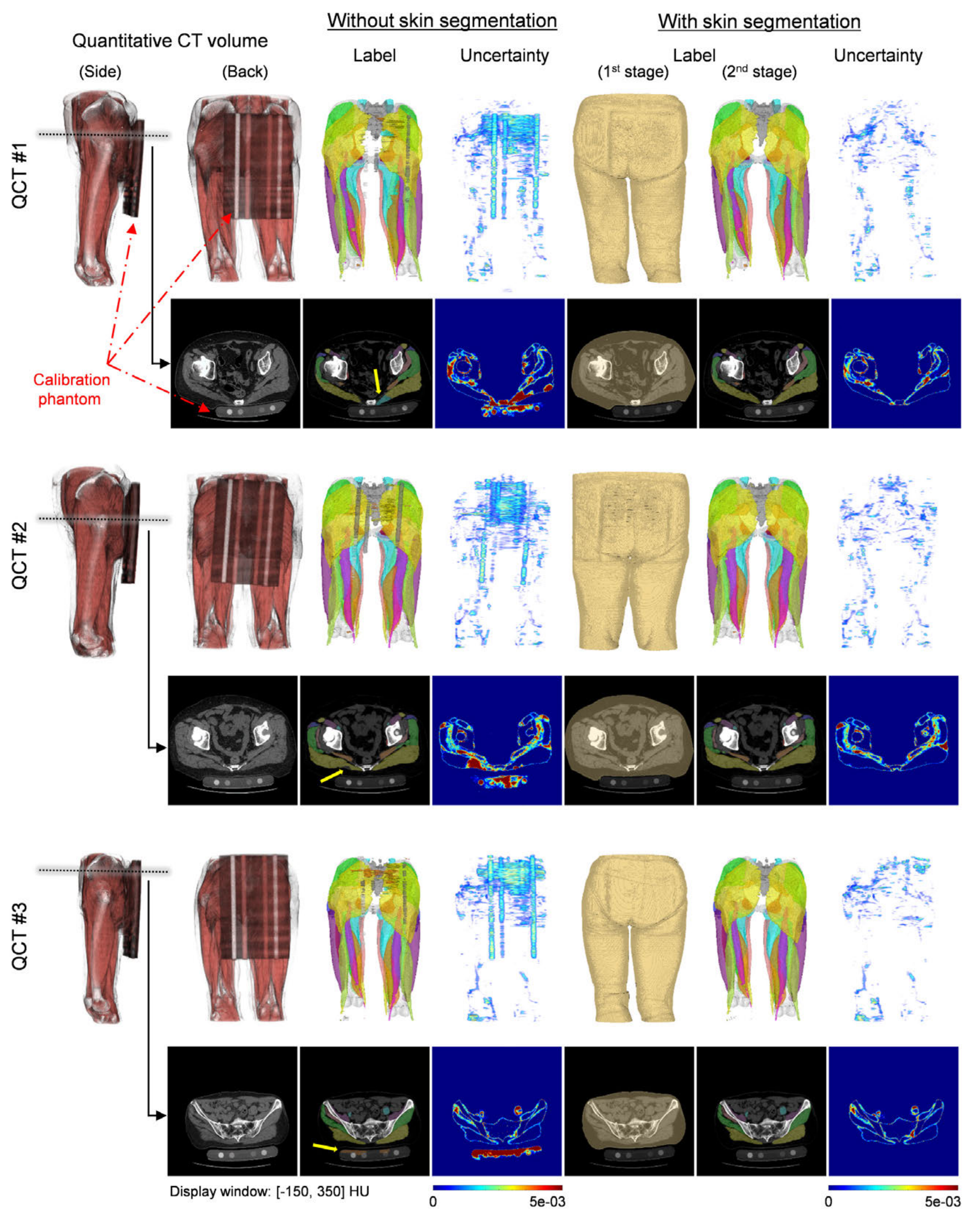}
			\caption{Qualitative evaluation of the effectiveness of the skin surface segmentation step in the quantitative CT (QCT) volumes, which scanned with an intensity calibration phantom placed near the skin surface. Bayesian U-Net was trained using 20 CT volumes in the THA data set without containing the calibration phantom. Three cases from our QCT data set (independent from the THA data set) were shown. Note that, when the skin segmentation step was not applied (3rd and 4th columns), the calibration phantom was wrongly segmented, indicated high uncertainty, and mis-segmentation in the muscles near the phantom boundary was observed (yellow arrows), while the skin segmentation step corrected these errors (5th, 6th, and 7th columns) and improved accuracy of the muscle segmentation.
			}
		\end{figure}

	\clearpage

	\subsection*{}
		\begin{figure}[!]
			\centering
			\includegraphics[width=0.65\textwidth]{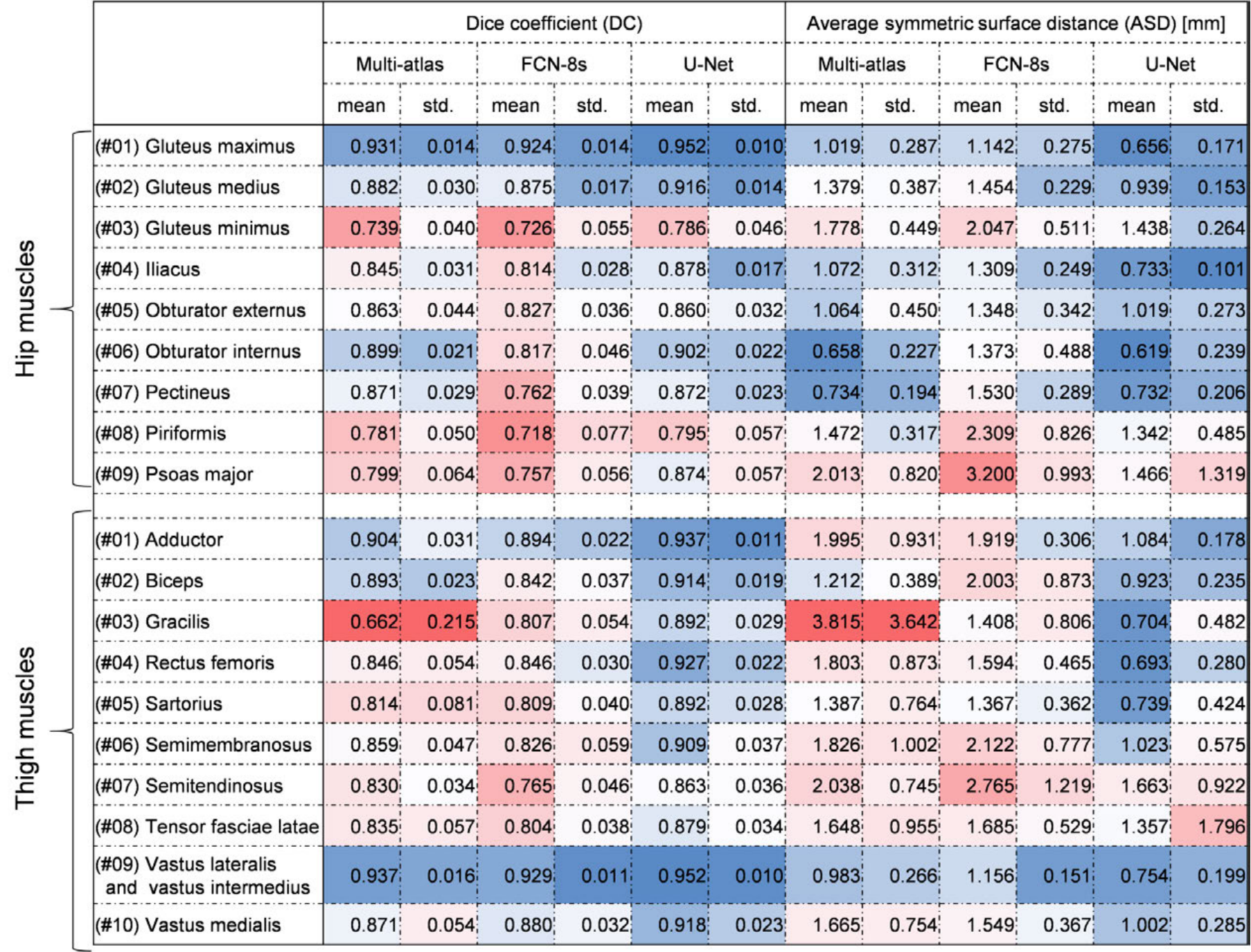}
			\caption{Accuracy of individual muscular structures for 20 patients in THA data set with the hierarchical multi-atlas method \cite{yokota2018automated}, FCN-8s and U-Net. Blue color shows high averaged accuracy or low variance.
			}
		\end{figure}

		\begin{figure}[!]
			\centering
			\includegraphics[width=0.78\textwidth]{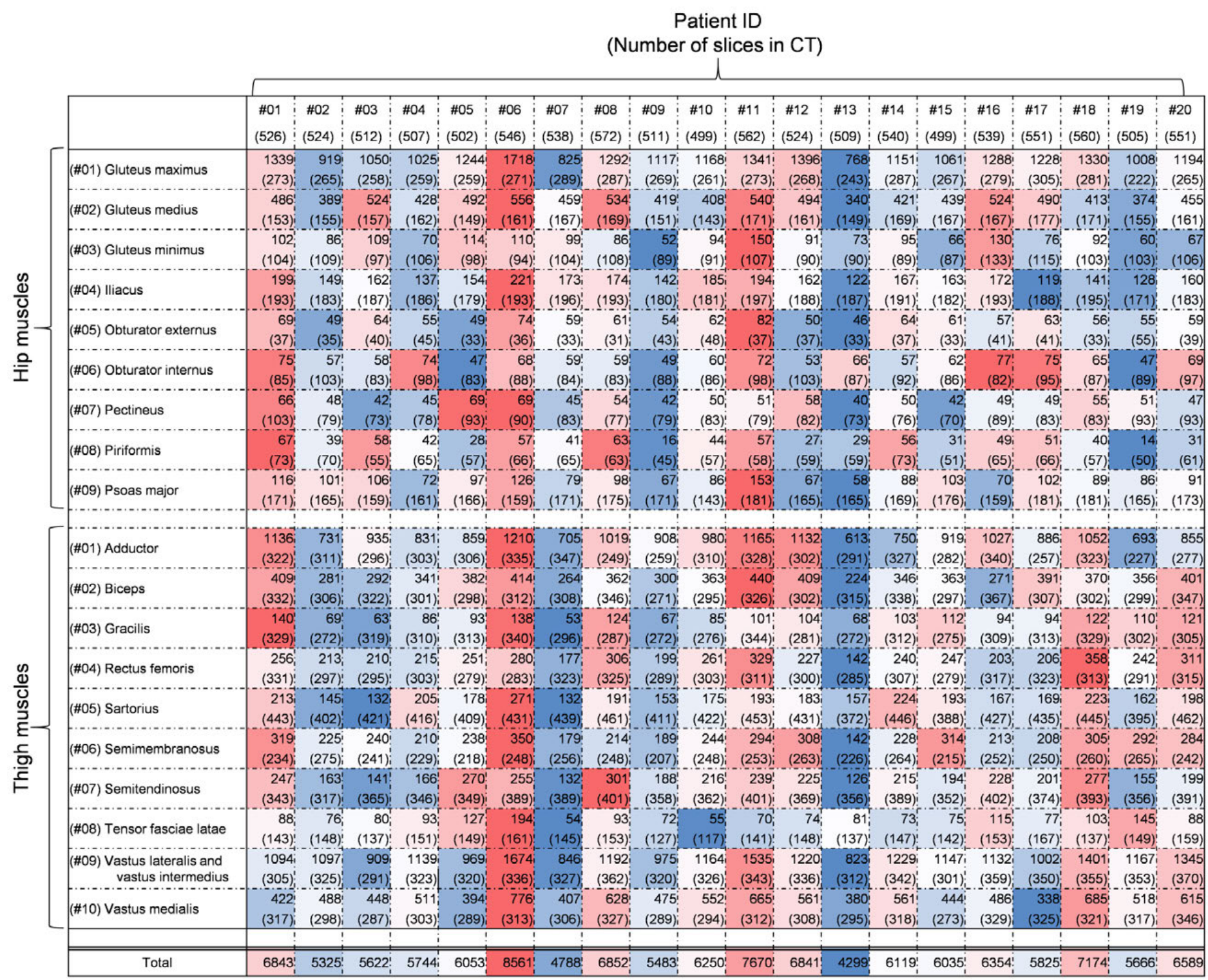}
			\caption{List of the volume in mL and the number of CT slices (in brackets) for individual muscular structures of each patient in THA data set. Color indicates the relative volume per structure (blue indicates small volume).
			}
		\end{figure}

\end{document}